# Pulse Sequences for Efficient Multi-Cycle Terahertz Generation in Periodically Poled Lithium Niobate

Koustuban Ravi[1,3*], Damian N. Schimpf[1] and Franz X. Kärtner[1,3]

[1] Center for Free-Electron Laser Science, Deutsches Elektronen Synchrotron, Hamburg 22607, Germany
[2] Department of Physics and the Hamburg Center for Ultrafast Imaging, University of Hamburg, Germany
[3] Department of Electrical Engineering and Computer Science, Research Laboratory of Electronics, Massachusetts Institute of Technology, Cambridge, MA 02139, USA

koust@mit.edu

**Abstract:** The use of laser pulse sequences to drive the cascaded difference frequency generation of high energy, high peak-power and multi-cycle terahertz pulses in cryogenically cooled periodically poled lithium niobate is proposed. Detailed simulations considering the coupled nonlinear interaction of terahertz and optical waves show that unprecedented optical-to-terahertz energy conversion efficiencies > 5%, peak electric fields of hundred(s) of Mega volts/meter at terahertz pulse durations of hundred(s) of picoseconds can be achieved. The proposed methods are shown to circumvent laser-induced damage at Joule-level pumping by 1µm lasers to enable multi-cycle terahertz sources with pulse energies >> 10 milli-joules. Various pulse sequence formats are proposed and analyzed. Numerical calculations for periodically poled structures accounting for cascaded difference frequency generation, self-phase-modulation, cascaded second harmonic generation and laser induced damage are introduced. Unprecedented studies of the physics governing terahertz generation in this high conversion efficiency regime, limitations and practical considerations are discussed. Varying the poling period along the crystal length and further reduction of absorption is shown to lead to even higher energy conversion efficiencies >>10%. An analytic formulation valid for arbitrary pulse formats and closed-form expressions for important cases are presented. Parameters optimizing conversion efficiency in the 0.1-1 THz range, the corresponding peak electric fields, crystal lengths and terahertz pulse properties are furnished.

## 1. Introduction

Multi-cycle or narrowband terahertz pulses in the frequency range of 0.1 to 1 THz have garnered interest as drivers of compact particle acceleration [1, 2], coherent X-ray generation [3] and electron beam diagnostics. An impediment to the widespread deployment of these applications has been the inadequate development of accessible sources of narrowband terahertz radiation (hundred(s) of picoseconds (ps) pulse duration) with *simultaneously* high pulse energy (> 10 milli-joules (mJ)) and peak powers (> 100 Mega-Volt per meter (MV/m) peak electric fields).

Among existing methods, photoconductive switches can be efficient [4] but relatively challenging to scale to high pulse energies, vacuum electronic devices [5] are limited in their frequency of operation and peak powers while free electron lasers [6] are relatively expensive.

With the rapid scaling of laser pulse energies, laser driven approaches employing second order nonlinear processes such as difference frequency generation (DFG) or optical rectification (OR) are promising. However, scaling this approach to high terahertz pulse energies will require achieving high optical-to-terahertz energy conversion efficiencies (or conversion efficiency for short) as well as the development of high energy optical lasers. Here, we describe approaches to improve conversion efficiencies for multi-cycle terahertz generation, which are still relatively low at the sub-percent level. This problem must be distinguished from broadband or single-cycle source development where percent level conversion efficiencies have been demonstrated [7, 8, 9]. Correspondingly, we only survey work pertinent to multi-cycle or narrowband sources.

Works using quasi-phase-matched (QPM) Gallium Arsenide (GaAs) [10, 11, 12, 13, 14] with conversion efficiencies of $10^{-4}$, Gallium Phosphide (GaP) [15] with conversion efficiencies of $10^{-6}$ have been reported. However, these materials require pumping by 1.3 µm or longer wavelengths where it is still relatively challenging to develop laser technology with the requisite pump pulse energies. Organic materials have produced multi-cycle radiation with $10^{-5}$ conversion efficiencies [16] but are also limited by the requirement of optical pumping at wavelengths of 1.3-2µm, large absorption coefficients at lower terahertz frequencies and laser-induced damage.

Lithium niobate (LN) possesses high second order nonlinearity and is compatible with rapidly developing 1µm [17] and established 800 nm laser technology. Furthermore, the large absorption in LN maybe reduced by cryogenic cooling. In combination with feasible Joule-class 1 µm lasers and large cross-section crystals, cryogenically cooled LN may thus offer solutions to the problem of high energy multi-cycle terahertz generation.

In LN, multi-cycle THz generation by interfering chirped and delayed copies of a pulse with tilted-pulse-fronts (TPF) has been demonstrated [18, 19]. However, TPFs have limitations induced by angular dispersion [20, 21, 22] which also affect other non-collinear approaches [23, 24] . Collinear geometries based on periodically poled lithium niobate (PPLN) [25] could circumvent angular dispersion based limitations but are characterized by a large walk-off between the optical and terahertz radiation. As a result, even the highest conversion efficiency for multi-cycle generation in PPLN crystals at cryogenic temperatures was only 0.1% [26].

For terahertz generation in the frequency range of 0.1 to 1 THz, the conversion efficiency will be limited to the sub-percent level even if every optical photon (300 THz) is converted to a terahertz photon, due to the large quantum defect. To surpass this limitation, repeated energy down-conversion of optical photons or cascaded DFG [27, 28, 29, 30] was proposed conceptually. However, methods to utilize this concept and produce large conversion efficiencies at the percent level or more for multi-cycle sources have not been demonstrated or even proposed.

Here, we study a family of approaches comprised of a sequence of pulses that can achieve very high conversion efficiencies > 5% in cryogenically cooled PPLN crystals. Employing a sequence of uniformly spaced pulses in time circumvents walk-off and coherently boosts the generated terahertz field. In combination with the low loss of cryogenically cooled PPLN crystals, this results in ~cm interaction lengths. Additionally, the low dispersion of LN in the 1µm region permits a large number of phase-matched, repeated energy down conversions of the optical pump photons. Finally, distributing the pump energy over a long sequence mitigates nonlinear phase accumulation and laser-induced damage. For Joule-level pumping, this reduces the required PPLN aperture area to 1-2 $cm^2$, which has already been demonstrated [31]. As a result, the set of proposed methods can produce the desired terahertz sources with pulse energies >>10 mJ and peak electric fields of hundreds of MV/m. It is worth pointing out that recently we proposed another set of approaches employing terahertz driven cascaded parametric amplification which yield similar performance [32].

We discuss different approaches to realize a sequence of pulses in time such as direct generation of a burst of pulses, beating multiple frequency lines and interfering chirped and delayed broadband pulses. Related pulse formats have been briefly explored in bulk ZnTe [28] [33], with TPFs in bulk LN at room temperature [19] and quasi-phase-matched GaAs [13, 11]. However, the work presented here is significantly different for a few reasons. Firstly, a different system, i.e. cryogenically cooled PPLN is studied. Secondly, here the emphasis is on feasible designs that enable dramatic cascaded DFG to achieve unprecedented conversion efficiencies >>1%, particularly for high energy pumping. Furthermore, unprecedented studies of the physics in this high conversion efficiency regime, limiting factors and corresponding correction mechanisms are presented. Finally, in this work, the various pulse sequence formats are treated on a common footing to elucidate the overarching mechanisms unifying their behavior.

Since such high conversion efficiencies can only be obtained by cascaded DFG, calculations incorporating pump depletion are imperative. In *Section 2*, a numerical model for PPLNs which considers the coupled nonlinear interaction of optical and THz fields (or pump depletion) is introduced. The model accounts for dispersion, absorption, cascaded DFG, self-phase-modulation (SPM), cascaded second-harmonic-generation (SHG) and laser induced damage. An analytic formulation considering absorption and dispersion based on undepleted pump approximations is provided for arbitrary pulse formats. At low conversion efficiencies, the analytic calculations show excellent agreement with the numerical calculations. However, the results deviate as conversion efficiencies of several percent are obtained. In *Section 3*, the general physics underpinning terahertz generation with pulse sequences is described. Cascaded DFG causes a red-shift and broadening of the optical spectrum (henceforth referred to collectively as cascading effects) which is seen to result in a modification of the phase-matching condition. This results in a drop in conversion efficiency. Varying the PPLN period along the propagation direction and further reduction of absorption by reducing the crystal temperature to 10K can thus lead to significantly higher conversion efficiencies. In *Section 4*, the conversion efficiencies, peak terahertz electric fields and terahertz pulse properties produced by various pulse sequence formats over a large parameter space are analyzed. Closed form expressions are provided for each format. In *Section 5*, various practical considerations such as self-focusing, THz diffraction, pump laser architectures and crystal fabrication are discussed.

## 2. Theory
### 2.1 Depleted formulation for QPM structures

Since the proposed set of methods is envisaged to operate in the several percent energy conversion efficiency regime where every optical photon yields multiple terahertz photons by cascaded DFG, it is necessary to simultaneously solve for both optical and terahertz spectral components as in Eqs. (1)-(2).

$$\frac{dA_{THz}(\Omega,z)}{dz} = -\frac{\alpha(\Omega)}{2} A_{THz}(\Omega,z) - \frac{j\Omega^2 \chi^{(2)}(z)}{2k(\Omega)c^2} e^{jk(\Omega)z} \int_0^\infty E_{op}(\omega+\Omega,z) E_{op}^*(\omega,z) d\omega \quad (1)$$

$$\frac{dA_{op}(\omega,z)}{dz} = -\frac{j\omega^2 \chi^{(2)}(z)}{2k(\omega)c^2} e^{jk(\omega)z} \left\{ \int_0^\infty E_{op}(\omega+\Omega,z) E_{THz}^*(\Omega,z) + E_{op}(\omega-\Omega,z) E_{THz}(\Omega,z) d\Omega \right\}$$
$$- \frac{j\varepsilon_0 \omega_0 n(\omega_0) n_{2,eff}}{2} \mathrm{F}\left\{ A_{op}(t,z) \left| A_{op}(t,z) \right|^2 \right\} \quad (2)$$

In Eq.(1), the spectral component $E_{THz}(\Omega,z)$ of the terahertz electric field at angular frequency $\Omega$ propagating in the positive $z$ direction is described by the complex scalar $A_{THz}(\Omega,z)e^{-jk(\Omega)z}$ where $A_{THz}(\Omega)$ represents the envelope and $k(\Omega)$ represents the wave number at $\Omega$ (See Appendix A for Fourier transform conventions). The optical laser field contains a number of spectral components $E_{op}(\omega,z) = A_{op}(\omega,z)e^{-jk(\omega)z}$. Here $A_{op}(\omega,z)$ represents the envelope at angular frequency $\omega$ and $k(\omega) = \omega n(\omega)/c$ is the corresponding wave number. The exact dispersion of the optical refractive index over a large bandwidth is accounted for by $n(\omega)$.

In Eq.(1), the first term on the right hand side (RHS) accounts for terahertz absorption. The second term on the RHS of Eq.(1) corresponds to the aggregate of all possible DFG processes between various spectral components of the optical field. The periodic inversion of the second order nonlinearity in PPLN crystals is explicitly accounted for by the $\chi^{(2)}(z)$ parameter in Eqs.(1)-(2). Explicitly accounting for the exact variation of $\chi^{(2)}(z)$ directly considers all orders of forward propagating QPM waves in the PPLN crystal. The specified pulse formats will only permit phase-matching of harmonics of a single terahertz frequency and hence backward phase-matched waves need not be considered.

Equation 2 represents the corresponding evolution of the envelope $A_{op}(\omega, z)$. The first term on the RHS of Eq.(1) represents the generation of optical radiation due to DFG between higher optical frequencies and THz radiation. This term is responsible for cascading effects. The second term corresponds to sum-frequency generation (SFG) between lower optical frequency components and terahertz radiation and causes some blue shift of the optical spectrum. The final term corresponds to the cumulative SPM effect which is a third order process.

While second harmonic generation (SHG) is highly phase mismatched in lithium niobate ($\Delta k \sim 10^6 \text{m}^{-1}$), phase-mismatched cascaded SHG (not to be confused with cascaded DFG) can influence the optical pump radiation as an effective third order effect [34]. The explicit consideration of SHG would significantly increase computation time. However, they may be accounted for by a cascaded SHG approximation absorbed into the effective nonlinear refractive index $n_{2,eff}$ term (See Appendix B). The optical-to-terahertz energy conversion efficiency (or conversion efficiency) $\eta$ is readily calculated by aggregating energy over all terahertz spectral components as follows. Here, $F_{pump}$ is the input optical pump fluence.

$$\eta(z) = \frac{\pi \varepsilon_0 c \int_0^\infty n(\Omega) |A(\Omega, z)|^2 d\Omega}{F_{pump}} \quad (3)$$

The model thus presented considers only one spatial dimension since transverse beam effects are not expected to be significant for the large pump beam sizes pertinent to this work. The length scales of transverse effects estimated in Section 5 justify this assumption.

**2.2 Analytic formulation for arbitrary pulse formats**
In addition to the above depleted calculations, analytic expressions for arbitrary pulse formats using undepleted pump approximations are derived. While they do not account for cascading effects, they show good qualitative and quantitative agreement with full numerical simulations considering cascading effects. These expressions alleviate the computational challenge of exploring the problem over large parametric spaces, provide overview and consistency checks.

We set $E_{op}(\omega, z) = A_{op}(\omega, z)e^{-jk(\omega)z}$ and retain only the first term in the Fourier series expansion of $\chi^{(2)}(z) = \sum_{m \neq 0} 2/(m\pi) \sin(m\pi/2) \chi_0^{(2)} e^{-j2m\pi\Lambda^{-1}z}$ in the second term on the RHS of Eq.(1). Furthermore, the optical wave numbers $k(\omega+\Omega), k(\omega)$ are expanded via a Taylor series around the central angular frequency $\omega_0$ of the pump to the second order to result in Eq.(4).

$$P_{THz}(\Omega) = \varepsilon_0 \chi_{eff}^{(2)} e^{-j\Delta k z} \int_0^\infty A_{op}(\omega+\Omega) A_{op}*(\omega) e^{-j\left(\beta''\Omega(\omega-\omega_0) + \frac{\beta''\Omega^2}{2}\right)z} d\omega \quad (4)$$

In Eq.(4), $\Delta k = \Omega[n(\Omega) - n_g]c^{-1} - 2\pi\Lambda^{-1}$ is the phase mismatch between the optical pump and generated THz radiation in a PPLN crystal. Here, $n_g$ is the optical group refractive index and $n(\Omega)$ is the THz phase refractive index at angular frequency $\Omega$. The parameter $\Lambda$ corresponds to the period of the PPLN crystal. The periodic inversion of the nonlinearity in the material imparts a momentum $2\pi\Lambda^{-1}$, which appears in the expression for phase-mismatch. Setting, $\Delta k = 0$, provides the required QPM period $\Lambda = c(f_{THz}[n(\Omega) - n_g])^{-1}$ for the phase-matched terahertz frequency $f_{THz}$ (not angular frequency).

In Eq.(4), $\chi_{eff}^{(2)}$ is the effective second order nonlinearity corresponding to the first term in the Fourier series expansion of $\chi^{(2)}(z)$ and $\beta''$ is the group velocity dispersion due to material dispersion (GVD-MD). Therefore, only the second order term in dispersion is accounted for in the analytic formulation which is contrary to Eqs.(1)-(2) where the complete dispersion of the optical refractive index is considered. Equation (4) is then integrated to yield the following analytic expression for the terahertz envelope (See Appendix A for derivation).

$$A_{THz}(\Omega, z) = \frac{-j\Omega \chi_{eff}^{(2)}}{n(\Omega)c(\alpha(\Omega) - 2j\Delta k)} \left( \mathbf{F}_{t\to\omega}\{A_{op}(t)A_{op}^*(t + \beta''\Omega z)\} e^{-j\left(\Delta k + \frac{\beta''\Omega^2}{2}\right)z} - \mathbf{F}_{t\to\omega}\{|A_{op}(t)|^2\} e^{-\frac{\alpha(\Omega)z}{2}} \right)$$
(5)

In Eq.(5), $\mathbf{F}_{t\to\omega}\{.\}$ corresponds to the Fourier transform between the time ($t$), and angular frequency ($\omega$) domains. Upon setting $\beta'' = 0$, we see that the $A_{THz}(\Omega, z)$ is directly proportional to $\mathbf{F}_{t\to\omega}(|A_{op}(t)|^2)$ or the Fourier transform of the optical intensity profile. Note that no assumption about the optical spectrum has been made in Eq.(3), i.e. it is valid for an arbitrary optical pump pulse format.

Equation (5) is verified by evaluating it for the case of a single unchirped Gaussian pulse in a lossless PPLN crystal without dispersion. The second order nonlinear coefficient $\chi_{eff}^{(2)} = 2d_{eff}$. The absolute value of the field component squared for this case is shown in Eq.(6). The expression is in agreement with other formulae in the literature [11, 35].

$$|A_{THz}(\Omega, z)|^2 = \frac{\Omega^2 d_{eff}^2 F_{pump}^2}{c^4 \pi^2 \varepsilon_0^2 n(\omega_0)^2 n^2(\Omega)} e^{-\frac{\Omega^2 \tau^2}{4}} \text{sinc}^2\left(\frac{z\Delta n}{c} \frac{\Omega - \Omega_{THz}}{2}\right) z^2$$
(6)

In Eq.(6), $\Omega_{THz} = 2\pi f_{THz}$ is the angular frequency of the phase-matched terahertz wave and $\Delta n = n(\Omega) - n_g$. Equation (6) illuminates certain important features. For instance, barring effects of absorption, the terahertz conversion efficiency would increase quadratically with terahertz frequency. If phase-matching is broadband, then the sinc$^2$ function is ~ 1 and the conversion efficiency (Eq.(3)) increases quadratically with crystal length $z$. However, if phase-matching is narrowband and there is large walk-off between the optical pump and generated terahertz wave (large $\Delta n$), as is the case for LN, the sinc$^2$ function in Eq.(6) will have a small linewidth centered about the phase-matched terahertz frequency $f_{THz}$ and may be approximated as a Dirac-Delta function $\delta(\Delta nz[\Omega - \Omega_{THz}]/c)$. Invoking this approximation leads the conversion efficiency $\eta$ from Eq.(3) to only increase linearly with crystal length $z$ in the absence of absorption. Thus, the walk-off limitation present in lithium niobate may be deduced from Eq.(6).

**3. Terahertz generation with pulse sequences: A general discussion**
In this paper, three pulse formats suitable for high energy terahertz generation using cryogenically cooled PPLNs are presented. They comprise of (i) a burst of pulses of equal intensity, (ii) a set of quasi-continuous wave (quasi-CW) lines and (iii) interfering chirped and delayed broadband pulses. All of these formats constitute different forms of pulse sequences in time with common underlying physics. In this section, we provide an overview of the overarching mechanisms of terahertz generation using pulse sequences, their limitations and corresponding correction mechanisms.

*3.1 Physical motivation: Alleviating walk-off and laser induced damage*

Consider the case of a single optical pump pulse propagating through a PPLN crystal phase-matched for terahertz generation at frequency $f_{THz}$. The electric field of the optical pump pulse is aligned with the extraordinary axis of the crystal for maximum nonlinearity. The optical group refractive index $n_g$ = 2.21 at 1030 nm, while the terahertz phase refractive index is $n(\Omega)$ ~ 5. In the case of such large $\Delta n = n(\Omega) - n_g$, the optical pump pulse walks off very rapidly from the generated terahertz radiation. As a result, rather than adding on top of the already generated terahertz radiation, the optical pump merely adds another cycle to the back of it. This produces only a linear growth of conversion efficiency with length, in accordance with the discussion following Eq.(6).

Secondly, notwithstanding SPM or self-focusing effects, the permissible peak intensity (and hence fluence) of the optical pump pulse is limited by laser-induced damage (or damage). Therefore, continuously increasing the intensity of the pulse is not a feasible solution to scaling conversion efficiencies. Equivalently, this damage limitation prohibits the use of very large pump energies for feasible crystal apertures.

Now consider a sequence of two optical pump pulses instead of just one incident on a cryogenically cooled PPLN with crystal temperature $T$ =100 K as simulated in Fig.1a. The PPLN period $\Lambda$=237.74 µm has a spatially dependent second order nonlinearity $\chi^{(2)}(z)$ as shown in Fig.1. The crystal parameters correspond to Magnesium Oxide (MgO) doped (5 % mol.) congruent lithium niobate (See Table.1 for list, Appendix A for details). Each pump pulse (blue) is Gaussian with a transform limited (TL) full-width at half-maximum (FWHM) duration $\tau$ = 400fs optimized for terahertz generation at $f_{THz} = 0.5$ THz (See Section 4.1). If the pump pulses are separated by a time interval $\Delta t$ corresponding to the time period of the phase-matched terahertz wave $T_0 = f_{THz}^{-1} = 2$ps (or separated by a distance $\Delta z = c n_g^{-1} f_{THz}^{-1}$ inside the crystal), the second pump pulse will coherently add to the terahertz electric field (red) generated by the first ultrafast pulse as shown in Fig.1b-c. In general, a sequence of $N$ pump pulses will serially boost the terahertz field generated by the first pump pulse in the sequence, thereby alleviating walk-off.

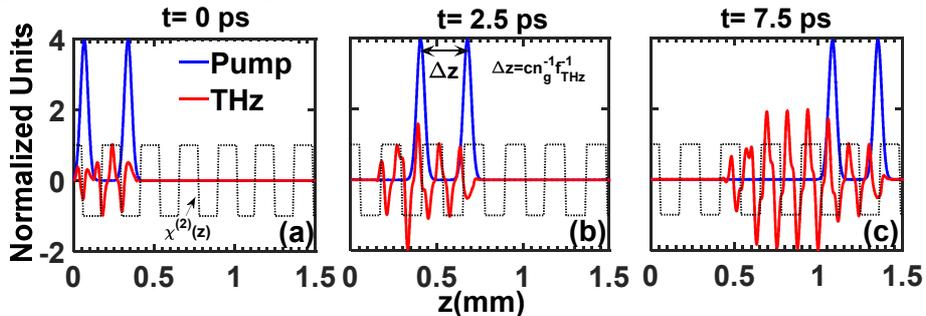

Fig.1 (a) A sequence of 2 optical pump pulses of 400 fs FWHM duration each separated by the time period $T_0$ (2 ps) corresponding to the inverse of the THz frequency (0.5 THz).(b)-(c) The second pulse coherently boosts the magnitude of the THz field generated by the first pulse.

The laser induced damage threshold intensity $I_d$ *reduces* as the square root of the pulse duration [36]. Consequently, the damage threshold fluence $F_d$ *increases* as the square root of the pulse duration. The lower limit of the reported values $F_d$ for a pulse duration of 10 ns in lithium niobate is ~ 1 GW/cm² [37]. Therefore, we use the following empirical expressions to determine $I_d$ and $F_d$. Among various factors, the quality of anti-reflection coatings, crystal growth methods, choice and concentration of dopants all influence laser-induced damage. For instance, in preliminary tests on Magnesium Oxide doped (5% mol.) congruent Lithium Niobate crystals, we recorded damage threshold values, four times larger than those in Eq.(7) [38].

$$I_d = 1(\tau_d/10ns)^{-1/2} \text{ GW/cm}^2 \quad (7a)$$

$$F_d = 10(\tau_d/10ns)^{1/2} \text{ J/cm}^2 \quad (7b)$$

A sequence of $N$ pump pulses may be considered to have an effective damage threshold pulse duration $\tau_d = N\tau$, since the time interval between pulses (~ps) is smaller compared to carrier decay time scales (>ns). Therefore, the permissible peak intensity scales as $N^{-1/2}$ but due to the serial reinforcement by $N$ pump pulses, the conversion efficiency increases as $N \times N^{-1/2}$ or as $\sqrt{N}$. Simultaneously, the damage fluence also scales as $\sqrt{N}$, which increases the energy loading capacity of the crystal. Thus, the approach of using a sequence of $N$ pump pulses alleviates walk-off, permits larger pump fluences and results in higher conversion efficiencies.

The scenario illustrated in Fig.1 addresses a case when every pump pulse in the sequence is of equal intensity. However, the above arguments are generally true even when this is not so. It is worth mentioning that the scaling of the damage threshold depends on the envelope of the pulse, which may result in quantitative alterations. The ramifications of pulse envelopes on damage shall be reported elsewhere.

*3.2 Terahertz generation mechanisms, limiting factors and correction mechanisms*

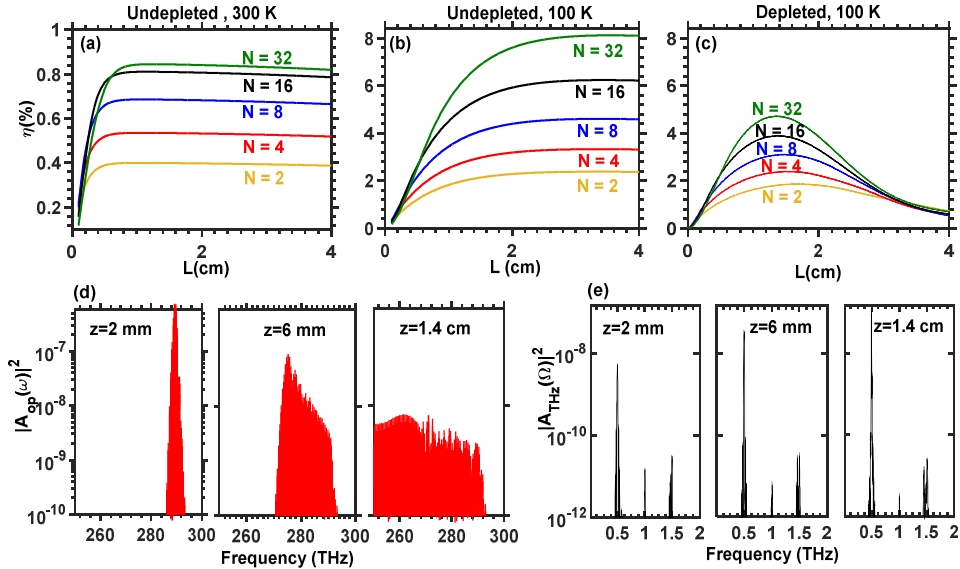

Fig.2 (a) Analytic undepleted calculations (Eq.5) at 300K for pulse trains with $N$ pulses. Large absorption results in low conversion efficiencies (b) Analytic calculations at 100 K showing enhanced conversion efficiencies due to increase in interaction lengths and reduction in absorption. (c)Depleted calculations (Eq.(1)-(2)) show good agreement with analytic calculations predicting >5% conversion efficiency at 100 K. Drop in conversion efficiency is due to a change in phase-matching produced by (d) spectral shift of optical pump. (e) Terahertz spectrum at 0.5 THz with large suppression of higher order quasi-phase-matched modes.

In Fig.2, we plot the conversion efficiency ($\eta$), according to Eq.(3), as a function of PPLN crystal length for a burst of $N$ optical pump pulses of equal intensity. As in Section 3.1, the simulated pump pulses have Gaussian full-width at half maximum (FWHM) durations $\tau =$

400 fs, separated by 2 ps each, optimized for generation of 0.5 THz radiation. The remaining material parameters are presented in Table.1.

In Fig.2a, undepleted calculations using Eq.5 at a crystal temperature $T$= 300 K are presented (PPLN period $\Lambda$=219 µm). From Fig.2a, we see that the optimal interaction lengths are < 0.5 cm due to large absorption (~7.5cm$^{-1}$ at 0.5 THz), which limits the conversion efficiency to < 1%. In Fig.2b, undepleted calculations at a crystal temperature of 100K ($\Lambda$=238 µm) are presented. The interaction lengths increase to ~ 2 cm due to a reduction in absorption (~1.36cm$^{-1}$ at 0.5 THz). The simultaneous increase in interaction length and reduction of absorption drastically enhances conversion efficiencies to 8%.

The values with larger $N$ initially increase at a slower rate since the intensity of a single pulse in the sequence is lower due to damage limitations (Eq.7a). However, due to mitigation of walk-off, they grow monotonically over a longer length, eventually resulting in higher conversion efficiencies. The optimal number of pulses in the sequence will be proportional to the optimal interaction lengths. Consequently, there is not as much benefit in increasing the number of pump pulses in the sequence from $N$=16 to 32 at $T$=300K as it is for $T$=100K.

Conversion efficiencies > 0.1%, are only possible when cascading is present, i.e. the optical pump undergoes repeated energy down conversions. Therefore, we evaluate the same cases via numerical solutions to Eqs.(1)-(2) in Fig.2c. These depleted pump calculations incorporating cascading effects agree quantitatively and qualitatively with that in Fig.2b during the initial increase of conversion efficiency. However, after reaching a smaller maximum value of ~ 5% at shorter interaction lengths ~1.5 cm, a fall in conversion efficiency, contrary to the saturation observed in analytic undepleted calculations (Fig 2a, 2b) is seen.

The fall in conversion efficiency is attributed to a change in the phase-matching condition caused by the spectral broadening and red-shift of the optical spectrum due to cascading. In Fig.2d, the optical spectrum as a function of crystal length is plotted for the $N$=32 case, which shows a steady red-shift due to repeated energy down conversion of the optical photons to terahertz photons. The corresponding terahertz spectrum (Fig.2e) is quasi-monochromatic at 0.5 THz, with a large suppression of higher harmonics.

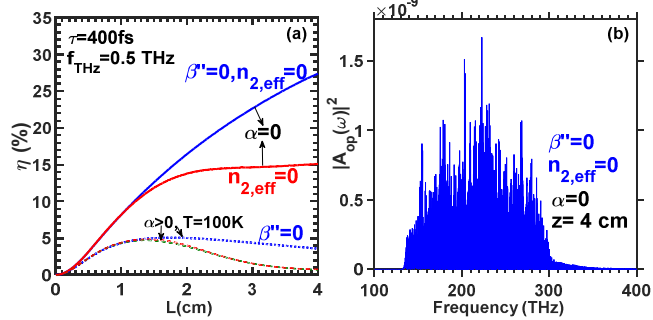

Fig.3a. Conversion efficiency as a function of crystal length with various effects selectively switched on or off for $N$=32 pulses. The drop in conversion efficiency observed is due to alteration of phase-matching conditions caused by spectral broadening and red-shift of the optical spectrum. Therefore, further abatement of losses and gradual variation of PPLN periods can yield conversion efficiencies >> 10% (blue,solid). (b) Optical spectrum for the case of a dispersion and absorption free medium (blue curve, Fig.3a) shows dramatic red-shift > 100 THz.

In Fig.3a, we perform simulations to identify the reasons for the drop in conversion efficiency observed in Fig.2c. Equations (1)-(2) are solved with various effects switched on or off to identify their relative influence. As in Fig.2, a burst of $N$=32 pulses of equal intensity, with FWHM durations $\tau$=400 fs each, separated by 2 ps are incident on a PPLN crystal phase-matched for 0.5 THz. When a medium with no dispersion (denoted by $\beta''$=0 in Fig.3, $n(\omega)$ is set to a constant value in Eqs.(1)-(2) ), no terahertz absorption ($\alpha$=0) and no SPM ($n_{2,eff}$=0) is assumed, the conversion efficiency even increases to ~ 30% (blue, solid curve). The inclusion of SPM does little to change this result. This suggests that further reduction of absorption by cooling to 10 K (0.25cm$^{-1}$ at 0.5 THz) and reduction of phase-slippage by gradually varying the PPLN period along the crystal length can lead to significantly higher conversion efficiencies.

However, when dispersion was switched on but absorption was zero (red, solid curve), the conversion efficiency was dispersion limited. In this case, the change in the phase-matching condition caused by cascading leads to a saturation of conversion efficiency. If absorption was switched on but dispersion was switched off (blue, dashed curve), phase-slippage is minimal but the conversion efficiency is limited by absorption. Consequently it saturates when the interaction length is on the order of the absorption length, similar to Fig.2b.

When both dispersion and absorption are switched on (red,dashed curve), the conversion efficiency drops rapidly after reaching a maximum value. This is because, after phase-matching is compromised and no net energy is transferred to the terahertz field, inclusion of absorption can only lead to an effective dissipation of THz energy. Further inclusion of SPM does not alter the results significantly (green,dashed curve overlaps with red,dashed curve). In Fig.3b, the optical spectrum corresponding to the case without absorption or dispersion (blue, solid curve in Fig.3a) is depicted. A dramatic spectral red shift and broadening is evident. The simulations in Fig.3, thus show that the drop in conversion efficiency observed in Fig.2c is caused by the change in phase-matching condition produced by the shift in the optical spectrum due to cascading. However, further mitigation of losses and reduction of phase-slippage can lead to significantly higher conversion efficiencies.

### 3.3 Temporal properties of terahertz pulses

Here, we examine the temporal properties of terahertz pulses generated by a sequence of optical pump pulses. First, the case of a burst of pump pulses with equal intensity is considered. The understanding is readily extended to other pulse sequence formats.

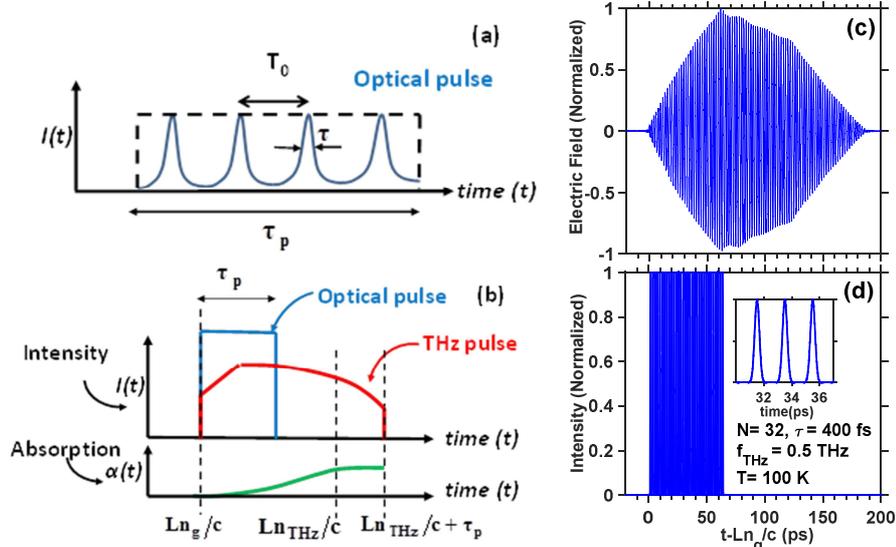

Fig.4a. Schematic of a burst of pump pulses of equal intensity. The total duration of the burst is $\tau_p$. (b) (Top panel Relative time scales of optical pump and generated THz pulses. The first terahertz to emerge is generated by the first optical pump pulse in the sequence near the exit of the crystal. The terahertz to emerge last is generated by the last pump pulse near the beginning of the crystal. (Bottom) Absorption is least for the terahertz generated near the exit and most for the terahertz generated at the beginning of the crystal. (c) Simulated terahertz field profiles for a burst of $N$=32, 400 fs (FWHM) Gaussian pulses. (d) Intensity of optical pump pulses

In Fig.4a, a schematic of a burst of optical pump pulses, separated by an interval $T_0 = f_{THz}^{-1}$ is depicted. The full-width at half-maximum duration of each pulse is $\tau$. The entire burst has a temporal extent $\tau_p$, which is the main quantity influencing the properties of the generated terahertz pulse. In Fig.4b, the relative time scales of the optical pump pulse and the generated terahertz pulse are depicted in the top panel. The bottom panel depicts the absorption experienced by the terahertz pulse over its duration. From Fig.4b, we see that the first pump pulse in the sequence entering the crystal at time $t=0$, emerges after a time $Ln_g c^{-1}$ and with it appears the first terahertz radiation generated near the crystal exit. The terahertz generated by the last pump pulse in the burst with duration $\tau_p$, at the beginning of the crystal, will emerge after a time $Ln_{THz}c^{-1}+\tau_p$ has elapsed. Therefore, the total temporal extent of the terahertz pulse is $L(n_{THz}-n_g)c^{-1}+\tau_p$. The shape of the terahertz pulse will be determined by overlap with the pump pulse sequence and absorption considerations as follows:

(i) The terahertz radiation emerging first, generated close to the exit of the crystal, is not serially boosted by the subsequent pump pulses in the sequence and will consequently have the smallest amplitude. Thereafter, the overlap of the generated terahertz with other pump pulses in the sequence linearly increases, up until the moment the final pump pulse emerges from the crystal, i.e. at $t=Ln_g c^{-1}+\tau_p$. After this time, the overlap saturates and begins to decline only after $t=Ln_{THz}c^{-1}$. This corresponds to the time when the terahertz radiation generated by the first optical pump pulse in the sequence emerges from the crystal. Naturally, any terahertz generated after this will overlap less with the pump pulse sequence.

(ii) The part of the terahertz pulse emerging first, generated close to the crystal exit experiences the least propagation distance and therefore least absorption. The terahertz generated by the first pump pulse in the sequence, at the beginning of the crystal, propagates through the entire crystal and hence experiences maximum absorption. It emerges after a time $Ln_{THz}c^{-1}$ has elapsed. Any terahertz radiation emerging after this time, experiences the same effective loss, since it too has propagated through the entire crystal.

Based on (i)-(ii), we understand the depicted terahertz pulse schematic in Fig.4b. It shows an initial increase in amplitude due to overlap and subsequent drop due to absorption. In Fig.4c-d, we plot the temporal waveforms of the terahertz electric field obtained from numerical simulations. The waveforms are plotted at the location of maximum conversion efficiency.

In Fig.4c, the terahertz electric field for a burst of $N=32$, $\tau=400$ fs pulses separated by 2 ps each are incident on a PPLN crystal phase-matched for 0.5 THz (Fig.2c) is shown. The conversion efficiency is maximized at $L=1.35$ cm (Fig.2c). The optical pump pulse sequence is depicted in Fig.4d. The inset in Fig.4d, shows the sequence of pulses separated by 2 ps each, corresponding to generation of 0.5 THz radiation. In line with previous reasoning, the amplitude of the terahertz electric field increases up until the duration of the pump pulse has elapsed. Thereafter, a decline initially due to increasing absorption is observed. A third steeper drop, corresponding to a decline in overlap occurs at about $t-Ln_g c^{-1}=$ 123 ps. This corresponds to the time $t=Ln_{THz}c^{-1}$ when the terahertz generated by the first pump pulse in the sequence at the beginning of the crystal has emerged.

In general, the terahertz electric field is proportional to $|A_{op}(t)|^2 \otimes \sin(2\pi f_{THz}t)$, or the convolution of the pump intensity envelope with a sinusoidal function at the frequency of the phase-matched terahertz wave. However, the critical time instants described in the overlap process above are similar for various pulse sequence formats.

### 3.4 Optimizing the number of pulses in a sequence

A recurring theme in the various pulse formats described will be the optimal number of pump pulses in the sequence or the optimal value of the effective pulse duration $\tau_p$ (See Fig.4a). An intuitive estimate of this is obtained by examining the time scales of the optical pump pulse sequence and generated THz pulses as viewed at the end of a crystal of length $L$ in Fig.4b. For most efficacious terahertz generation, temporal overlap between optical and terahertz pulses must be maximized. This translates to the condition $\tau_p > L(n_{THz} - n_g)c^{-1}$.

However due to absorption delineated in Fig.4b, a trade-off exists. Since the absorption saturates after $t = Ln_{THz}c^{-1}$, the fraction of time over which absorption saturation is present must be minimized. Therefore, one would desire $\tau_p < L(n_{THz} - n_g)c^{-1}$. Considering both constraints, an estimate for the optimal effective pulse length may be $\tau_p \sim L(n_{THz} - n_g)c^{-1}$.

The maximum crystal length $L = L_{max}$ itself is set by the strongest limitation to terahertz generation. This approximate condition on the value of $\tau_p$ agrees well with various simulations from ensuing sections. Quantitative corrections to this picture appear upon considering details of the pump pulse envelope.

### 3.5 Simulation methods and parameters

All ensuing calculations assume MgO doped (5% mol.) congruent lithium niobate as the nonlinear material. The crystal temperature is fixed at 100K. Material and other parameters used in the calculations are tabulated below.

| Parameter | Analytic calculations | Numerical calculations |
|---|---|---|
| Second order susceptibility (T=100 and T=300 K) | $\chi^{(2)}_{eff} = 2\pi^{-1}\chi^{(2)}_0$ $\chi^{(2)}_0$ =336 pm/V | $\chi^{(2)}(z) = \chi^{(2)}_0 \mathbf{sgn}\{\cos(2\pi\Lambda^{-1}z)\}$ $\chi^{(2)}_0$ =336 pm/V , $\Lambda$ (PPLN period, Appendix.A) |
| Central pump wavelength | $\lambda_0$=1030 nm | $\lambda_0$=1030 nm |
| Nonlinear refractive index | - | $n_{2,eff}$=1.25×10$^{-19}$ W/m$^2$ [34] (Appendix A) |
| Total fluence | $F_d = 10(\tau_d/10ns)^{0.5}$ Jcm$^{-2}$ | $F_d = 10(\tau_d/10ns)^{0.5}$ Jcm$^{-2}$ |
| Terahertz Refractive indices and absorption coefficients(Temperature dependent) | [39], [40]Appendix A) | [39], [40] (Appendix A) |
| Optical refractive indices(Temperature dependent) | - | [41] |
| Group velocity dispersion | β"=5×10$^5$ fs$^2$/m | - |
| Numerical method | - | 4$^{th}$ order Runge-Kutta |
| Range of crystal lengths | 0 to 5 cm | 0 to 5 cm |
| Simulated Frequency Bandwidth | 40 THz | 300 THz |
| Frequency resolution | 0.2 GHz | 0.5 GHz |
| Spatial resolution | - | 1.25µm |

Table.1. Parameters used for analytic and numerical solutions

The large cascading of the optical spectrum and long terahertz pulses necessitate simulations with large time-bandwidth products. The accuracy of simulations was verified by examining the numerical conservation of total energy (Appendix B.2). A bandwidth of 300 THz, a frequency resolution of 0.5 GHz and spatial resolution of 1.25µm is necessary. For available large aperture PPLN lengths of ≤ 5 cm, this translates to computational domain sizes with dimensions of 600,000×40,000 or 24 billion points.

The direct optimization of the numerical solution to Eqs (1)-(2) is thus impractical. Fast analytic calculations based on Eq.(5) are therefore used to obtain optimal pump parameters for maximizing conversion efficiency. Full numerical solutions are then performed to obtain accurate values of (i) conversion efficiency, (ii) peak electric field, (iii) terahertz pulse durations and (iv) optimal crystal lengths corresponding to these optimal pump parameters.

**4 Results**
**4.1 Burst pulse format**
We further analyze terahertz generation due to a burst of $N$ pump pulses with equal intensity. The electric field envelope of such a train is given by $A_{op}(t) = \sum_{n=1}^{N} E_0 e^{\frac{-2\ln 2(t-nT_0)^2}{\tau^2}}$. Here, $\tau$ is the FWHM pulse duration of each pulse. The pulses are separated from each other by a time $T_0 = f_{THz}^{-1}$. The FWHM duration $\tau = (2\ln 2)^{1/2} \tau_{e^{-2}}$, where $\tau_{e^{-2}}$ is the $e^{-2}$ pulse duration. Using Eq.(5), we obtain the following closed form expression for the spectral component $A_{THz}(\Omega, z)$ in the undepleted limit.

$$A_{THz}(\Omega, z) = \frac{-j\Omega \chi_{eff}^{(2)} (F_{pump}/N) e^{\frac{j(N-1)\Omega T_0}{2}}}{\pi c^2 \varepsilon_0 n(\omega_0) n(\Omega)(\alpha(\Omega) - 2j\Delta k)} \frac{\sin\left(\frac{N\Omega T_0}{2}\right)}{\sin\left(\frac{\Omega T_0}{2}\right)} \left( e^{-\frac{\Omega^2 \tau^2}{16\ln 2}\left(1 + \frac{4(2\ln 2)^2 \beta''^2 z^2}{\tau^4}\right)} e^{-j\Delta kz} - e^{-\frac{\Omega^2 \tau^2}{16\ln 2}} e^{-\frac{\alpha(\Omega)z}{2}} \right)$$

(8)

Inspecting Eq.(8), the optimal value of the pre-factor $\sin\left(\frac{N\Omega T_0}{2}\right)\left[\sin\left(\frac{\Omega T_0}{2}\right)\right]^{-1} = N$ for $T_0 = m f_{THz}^{-1}$, or for integral multiples of the time period of the phase-matched terahertz wave as was illustrated in Section 3.1. Assuming this condition, it is evident from Eq.(8) that the conversion efficiency scales as $\sqrt{N}$ since the total pump fluence $F_{pump}$ scales as $\sqrt{N}$ due to damage limitations. From the first exponential term within brackets, we see how the GVD-MD term $\beta''$ results in the spreading of the pump pulse in time. The second term within brackets corresponds to terahertz absorption. For $\beta'' \neq 0$, the conversion efficiency gradually vanishes due to reduction in peak intensity by spreading of the pulse in time, albeit rather slowly for small $\beta''$ (e.g. in Table.1). For example, a $\tau= 400$ fs pulse only suffers a pulse duration increase of less than 5 fs over 2 cm of propagation in lithium niobate. Furthermore, the pulse-spreading effect is less prevalent when the optical spectrum of the pulse is continuously increasing, as is the case with the numerical simulations based on Eqs.(1)-(2).

In Fig.5a, we plot the conversion efficiency for various $\tau$ and number of pump pulses $N$ in the sequence using Eq.(8) and simulation parameters from Table.1. The solid lines in Fig.5a represent the case of $N=2$. For small pump pulse durations (i.e. large bandwidth), the dominant limitation is the low pump fluence due to damage limitations. On the other hand, for large $\tau$, there is insufficient bandwidth for efficient terahertz generation. Consequently, an optimal $\tau$ is observed which decreases for larger terahertz frequencies. The dashed lines represent the case for $N=4$ pulses. The expectation of $\sqrt{N}$ scaling is evident upon comparing the two cases.

In Fig.5b, we plot the conversion efficiency as a function of $N$ using Eq.(8) at optimal values of $\tau$ for various $f_{THz}$. After the initial $\sqrt{N}$ scaling, at larger values of $N$, a saturation of conversion efficiency is observed. This is understood by recognizing that the parameter $NT_0$ assumes the role of the effective pump pulse duration $\tau_p$ described in Section 3.3-3.4. Since it was deduced that optimal $\tau_p \sim L(n_{THz} - n_g)c^{-1}$, the values of $N$ at which saturation of conversion efficiency occurs are expected to be roughly $L(n_{THz} - n_g)c^{-1}/T_0$. In the undepleted limit, the conversion efficiency begins to saturate at interaction lengths of $\sim 2$ cm at 0.5 THz

(See Fig.2b). This translates to a saturation of conversion efficiency for $N\sim100$ pulses which is well supported by the calculations in Fig.5b.

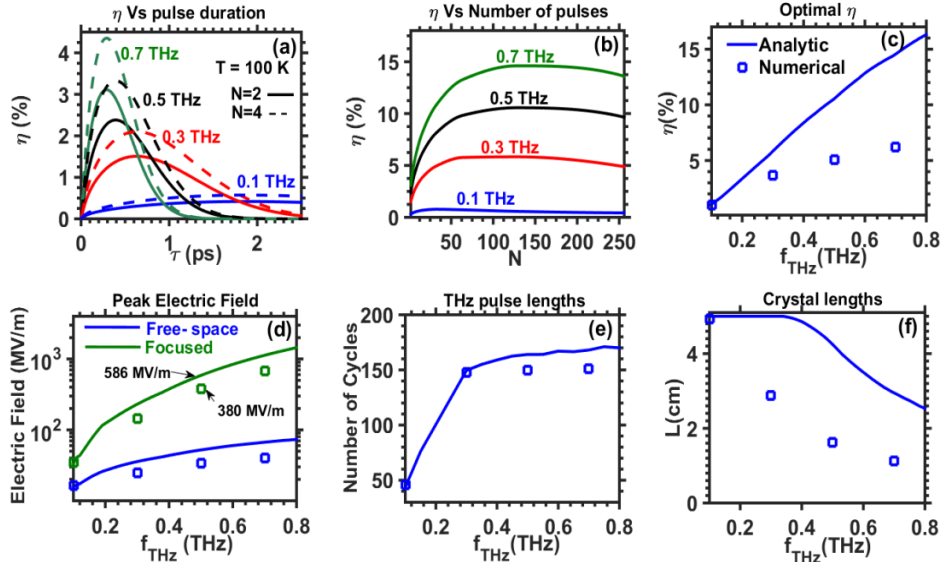

Fig.5 (a) Conversion efficiency ($\eta$) as a function of transform limited FWHM pulse duration $\tau$ at T=100K. Optimal $\tau$ values are inversely proportional to the generated frequency $f_{THz}$ (b) $\eta$ as a function of number of pulses $N$ in the sequence shows initial $\sqrt{N}$ scaling and eventual saturation when the effective pulse length $c\tau_p=cN/f_{THz}$ exceeds the walk-off length. (c) Optimal $\eta$ for various terahertz frequencies and corresponding calculations considering cascading (square markers). $\eta>5\%$ are predicted. (d) Peak free-space (blue) and focused (green) electric fields obtained analytically (solid) and numerically (square markers) predict field strengths of several hundred MV/m. (e) Number of cycles for various terahertz frequencies.(f) Optimal interaction lengths $L$ obtained analytically(solid) and numerically(square markers). Cascading effects modify phase-matching conditions and reduce interaction lengths in numerical calculations.

Figure 5c depicts the maximum conversion efficiency obtained by optimizing undepleted calculations based on Eq.(8) over $\tau$, $N$ and crystal length $L$ for various $f_{THz}$ (blue, solid). For optimal parameters, numerical solutions (square markers) to Eqs.(1)-(2) (i.e. depleted calculations accounting for cascading effects and SPM) are also performed. In general, an increase in conversion efficiency with frequency is observed in the presented frequency range.

The scaling is however not quadratic as simplistic expectations from Eq.(6),(8) suggest due to trade-offs incurred by terahertz absorption. The conversion efficiency may be expected to saturate at even higher terahertz frequencies due to the increasing values of absorption. For lower conversion efficiencies or smaller terahertz frequencies, almost exact quantitative agreement between analytic and numerical calculations is observed. However, at larger terahertz frequencies, greater deviation is observed due to a greater impact of the modification of phase-matching conditions by cascading effects. As described in Section 3, this results in a reduction of the optimal interaction lengths and conversion efficiencies. The properties of the emergent terahertz pulse are also altered. Despite these limitations, conversion efficiencies *in excess of 5%* at 0.5 THz are still predicted by numerical calculations, which is significantly higher in relation to existing approaches.

In Fig.5d, the peak free-space (blue) and focused (green) terahertz electric fields obtained from analytic (solid) and numerical simulations (square markers) are presented. The focused terahertz fields are estimated by scaling the free space values by a factor $w_{in}f_{THz}/c$, where

$w_{in}$ is the input beam radius of a pump laser pulse with a total energy of 1J and a flat-top (or top-hat) beam profile. Thus, $w_{in} = 1/\sqrt{\pi F_d}$, where $F_d$ is the damage threshold fluence.

The scaling reflects the reduction in terahertz beam radius from $w_{in}$ to $\sim cf_{THz}^{-1}$ (or the terahertz wavelength as may be obtained in linear accelerator structures [1]) upon focusing. Focused terahertz fields approaching the GV/m range are thus obtained using numerical calculations. Consistent with the case of conversion efficiencies in Fig.5c, the numerical calculations for corresponding peak terahertz electric fields are smaller compared to analytically obtained values at larger terahertz frequencies.

In Fig.5e, the numbers of terahertz field cycles corresponding to $e^{-2}$ pulse durations (full-width) are presented using analytic and numerical calculations. Naturally, lower terahertz frequencies contain smaller number of cycles, due to their longer time period. At 0.5 THz, ~100 cycles or 200 ps terahertz pulses are predicted. While the total temporal extent of terahertz pulses is altered due to the modification of interaction lengths in the case of numerical solutions, due to overlap effects, the $e^{-2}$ terahertz pulse duration remains roughly similar to analytic calculations.

In Fig.5f, the optimal interaction lengths obtained from numerical and analytic calculations are plotted. Quantitative agreement at 0.1 THz is obtained, which is consistent with trends from Figs.5c-5e. For larger terahertz frequencies, the modification of phase-matching conditions by cascading leads to shorter interaction lengths for the case of numerical calculations. The optimal crystal lengths obtained above are accessible with current PPLN technology.

**4.2 Difference frequency generation with multiple lines**

Beating two or more quasi-CW lines separated by the terahertz frequency $f_{THz}$ with each line corresponding to a pulse with transform limited duration of hundreds of picoseconds also generates a sequence of pulses separated by a time interval $T_0 = f_{THz}^{-1}$. The total spectrum of $M$ quasi-CW lines, where each line corresponds to a FWHM pulse duration $\tau$ is given by $A_{op}(\omega) = \sum_{m=0}^{M-1}\sqrt{\pi/4\ln 2}\, E_0 \tau e^{-\frac{(\omega-\omega_0-2\pi m f_{THz})^2\tau^2}{8\ln 2}}$. In time, $A_{op}(t)$ takes the form of a sequence of 'sub-pulses', each separated by $T_0$ with peak intensities following a Gaussian envelope (See Appendix B.3). As the number of lines $M$ increase, the duration of the individual 'sub-pulses' decrease and the peak intensity increases. However, the sub-pulses are still contained within an overall Gaussian envelope of FWHM duration $\tau$. This impacts the damage threshold intensity and fluence. For a set of $M$ lines, it can be shown that the relevant damage threshold duration $\tau_d = \tau/M$ (Appendix B.3). Upon inserting $A_{op}(\omega)$ into Eq.(5), we obtain Eq.(9) in the undepleted limit.

$$A_{THz}(\Omega, z) = \frac{-j\Omega\left(F_{pump}M^{-1}\right)e^{-\frac{(\Omega-2\pi f_{THz})^2\tau^2}{16\ln 2}}}{\pi c^2 \varepsilon_0 n(\omega_0)n(\Omega)(\alpha(\Omega)-2j\Delta k)}\left[\frac{\sin(\pi(M-1)\beta''\Omega f_{THz}z)}{\sin(\pi\beta''\Omega f_{THz}z)}e^{-\frac{2\ln 2\Omega^2\beta''^2 z^2}{2\tau^2}}e^{-j(\Delta k+\pi(M-1)\beta''\Omega f_{THz})z}-e^{-\frac{\alpha z}{2}}\right] \quad (9)$$

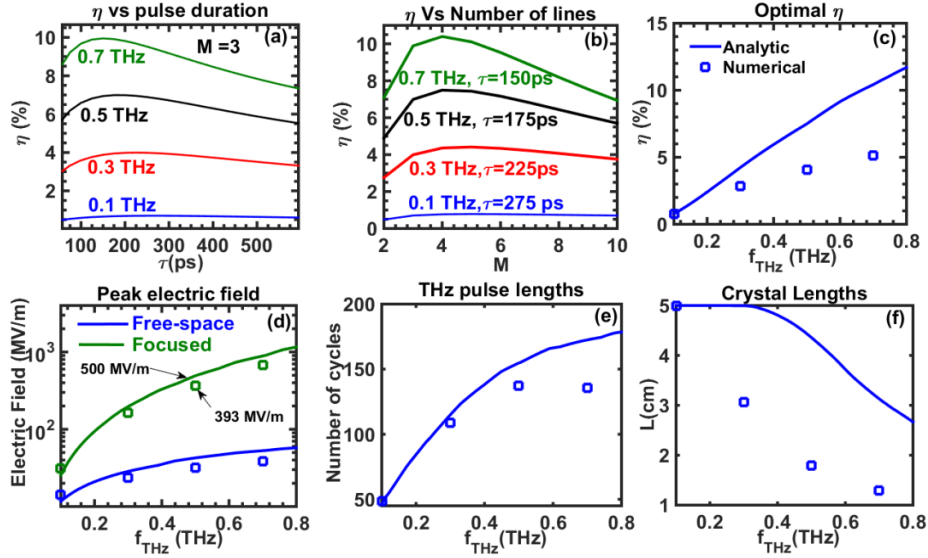

Fig.6(a).Conversion efficiency $\eta$ as a function of transform limited FWHM duration of each line $\tau$, showing that optimal pulse lengths $c\tau$ comparable to the walk-off length.(b) $\eta$ vs number of lines $M$, showing optimal $M\leq 5$. (c) Optimal $\eta$ obtained analytically (solid) and numerically (squares). Numerical simulations predict $\eta \sim 5\%$ at 0.5 THz. Deviations from analysis are due to modification of phase-matching conditions by cascading effects. (d) Peak free-space (blue) and focused (green) electric fields obtained analytically (solid) and numerically (squares) depicting field values of several hundred MV/m.(e) Terahertz pulses with hundreds of cycles are obtained and (f) reduced crystal lengths are predicted numerically .

The fluence per line is $M^{-1}F_{pump}$, where $F_{pump}$ is the total fluence bound by damage (Eq.7b) corresponding to $\tau_d = \tau/M$, i.e. $F_d = 10(\tau/(M.10ns))^{1/2}$. The exponential term outside the brackets in Eq.(9) shows the terahertz field centered about $f_{THz}$. The first exponential term inside the bracket of Eq.(9) represents the spreading of the pump pulses in time by GVD-MD. Note that the pre-factor $\sin(\pi(M-1)\beta''\Omega f_{THz} z)\sin(\pi\beta''\Omega f_{THz} z)^{-1}$ is exactly equal to $M-1$ for $\beta''=0$. This represents the situation where DFG processes between every pair of adjacent lines add up coherently. However, this may be modified for large values of $M$, $f_{THz}$ or $\beta''$. The second term in Eq.(9) corresponds to terahertz absorption. In the optimal case, the conversion efficiency which is proportional to $|A_{THz}(\Omega,z)|^2 F_{pump}^{-1}$ scales as $M^{-5/2}(M-1)^2$, for $F_{pump}=F_d$. This yields the optimal condition of $M\leq 5$.

Analytic calculations using Eq.(9) in Fig.6a depict the conversion efficiency as a function of $\tau$ for various $f_{THz}$ at $T=100K$ for $M=3$. In this case, $\tau$ takes on the role of the effective pump pulse duration $\tau_p$ introduced in Section 3.4. Thus, optimal values of $\tau$ are approximately equal to $L(n_{THz}-n_g)c^{-1}$. For example, at 0.5 THz, an optimal $\tau \sim 200$ ps is obtained which unsurprisingly resembles the optimal values of $NT_0$ for the burst pulse format ($N\sim 100$) in Section 4.1. In Fig.6b, the conversion efficiency for various $f_{THz}$ as a function of the number of lines $M$ is plotted using Eq.(9). As deduced in the preceding paragraph, the optimal number of lines is $M \leq 5$ for all frequencies.

In Fig.6c, the optimal conversion efficiency for various $f_{THz}$ are calculated analytically Eq.(9) (solid lines) and numerically using Eqs.(1)-(2) (square markers). Conversion efficiencies of $> 5\%$ at 0.5 THz are predicted by the numerical calculations including cascading and SPM. As with the previous case, the deviation of numerical from analytic results for larger terahertz frequencies or conversion efficiencies is owed to the greater impact of the modification of phase-matching conditions by cascading effects. In Fig.6d, the free-space (blue) and focused terahertz electric fields (green) are shown analytically (solid) and numerically (square markers). Numerical calculations reveal peak focused electric fields of several hundred MV/m. The lower conversion efficiencies obtained numerically in relation to analytic results yields proportionally smaller peak electric fields. The corresponding number of terahertz cycles obtained analytically and numerically are shown in Fig.6e. Due to overlap effects as in the case of Fig.5e, the number of cycles remains roughly similar for analytic and numerical calculations. However, in contrast to Section 4.1, due to the Gaussian envelope of the sequence of sub-pulses, the generated terahertz pulse also has a roughly Gaussian shape. As with the burst pulse format, the optimal interaction lengths obtained numerically are smaller compared to those obtained analytically for larger terahertz frequencies due to a modification of phase-matching conditions by cascading effects as evident in Fig.6f.

**4.3 Chirp and delay**

Interfering copies of chirped broadband pulses delayed with respect to each other offers an alternate method of generating a sequence of pulses. Since the bandwidth of the pulse is spread over time, the approach is similar to DFG between two long narrowband pulses. In particular, the method is attractive for use with off-the-shelf high energy broadband 800 nm titanium: sapphire (Ti:Sa) lasers. Here, we calculate the efficiency of the approach for cryogenically cooled PPLNs. We consider the optical laser field due to a chirp and delay setup as $A_{op}(t) = 2^{-1/2}\left( E_0 e^{-\frac{2\ln 2 t^2}{\tau^2}} e^{jbt^2} + E_0 e^{-\frac{2\ln 2 (t+\Delta t)^2}{\tau^2}} e^{jb(t+\Delta t)^2} \right)$. Here, $\tau$ is the FWHM chirped pulse duration and is related to the FWHM transform limited pulse duration by the chirp rate $b = 2\ln 2 (\tau \tau_{TL})^{-1}$. Inserting $A_{op}(t)$ into Eq. (5), we obtain the expression for the terahertz spectral component $A_{THz}(\Omega, z)$ as follows.

$$A_{THz}(\Omega, z) = \frac{-j\Omega(F_{pump}/2) e^{\frac{j\Omega\Delta t}{2}}}{\pi \varepsilon_0^2 c^2 n(\omega_0) n(\Omega)(\alpha(\Omega) - 2j\Delta k)} \left[ P_{THz}(\Omega, \Delta t') e^{-j\Delta k z} - P_{THz}(\Omega, \Delta t) e^{-\frac{\alpha z}{2}} \right] \quad (10a)$$

$$P_{THz}(\Omega, \Delta t) = \varepsilon_0 \chi_{eff}^{(2)} e^{-\frac{(\Omega - 2b\Delta t)^2 \tau^2}{16 \ln 2}} e^{-\frac{2 \ln 2 \Delta t^2}{2\tau^2}} \quad (10b)$$

$$\Delta t' = \Delta t - \beta'' \Omega z \quad (10c)$$

The total pump fluence $F_{pump}$ is bound by damage according to Eq.(7b). The damage threshold duration is equal to that of two narrowband pulses or the case of $M=2$ from Section 4.2 and is given by $\tau_d = \tau/2$ (See Appendix.B.3). Equation (10b) indicates that the generated terahertz is centered about the terahertz frequency $f_{THz} = b\Delta t/\pi$. Thus, the center frequency can be tuned with an appropriate chirp rate $b$ and delay $\Delta t$. However, a large delay causes $P_{THz}(\Omega, \Delta t)$ to drop due to insufficient overlap as evident in the exponential dependence on $\Delta t^2$ in Eq. (10b).

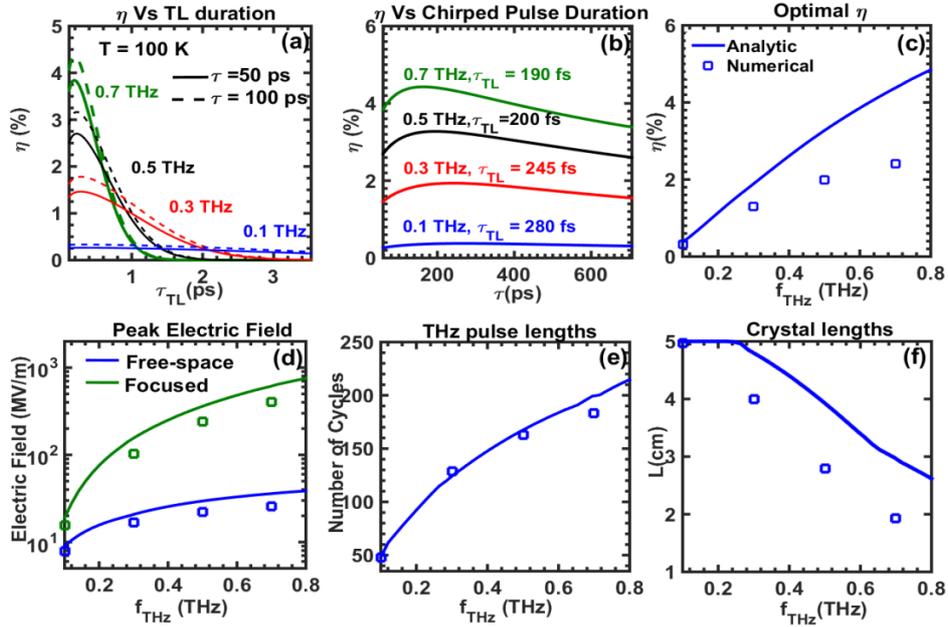

Fig.7a. Conversion efficiency η as a function of transform limited FWHM pulse duration $\tau_{TL}$. Low values are limited by dispersion (Eq.10c) and large values have insufficient bandwidth.(b) Optimal pulse length $c\tau$ is close to the walk-off length and is smaller for larger frequencies due to more absorption. (c) Optimal η obtained analytically (solid) and numerically (squares). η ~ 2% is predicted using numerical simulations. (d) Focused electric field gradient in the range of few hundred MV/m are predicted which is in close agreement with numerical results. (e) Terahertz pulse lengths corresponding to hundreds of cycles are obtained (f) Optimal crystal lengths predicted by numerical calculations are shorter .

In Fig.7a, the conversion efficiency $\eta$ is plotted for various transform limited FWHM pulse durations $\tau_{TL}$ and terahertz frequencies $f_{THz}$ at $T$=100K. At low values of $\tau_{TL}$, the GVD-MD term in Eq.(10c) leads to lower values of $\eta$. At large values of $\tau_{TL}$, there is insufficient bandwidth for efficient terahertz generation.  The dotted lines correspond to a $\tau$=50 ps, whereas the solid lines correspond $\tau$=100 ps. The similarity of their optima indicates that $\tau_{TL}$ and $\tau$ maybe treated as independent variables. In Fig.7b, we plot the conversion efficiency as a function of the chirped pulse duration $\tau$ at the optimum bandwidth values obtained from Fig.7a. In this case, $\tau$ assumes the form of the effective pulse duration $\tau_p$. Similar to the burst pulse format and the case of $M$ quasi-CW lines, the optimal values of $\tau$ also closely resemble the value $L(n_{THz} - n_g)c^{-1}$ as deduced in section 3.4.

For example, in Fig.7b, the optimal chirped pulse duration $\tau$= 200 ps at 0.5 THz which is similar to the values obtained for both the $M$ quasi-CW line case (Fig.6a), as well as the burst pulse case (Fig.5b). The conversion efficiencies obtained with the chirp and delay approach will be similar but lesser than the case with $M$= 2 quasi-CW lines (See Fig.6b,7b) due to the $e^{-\ln 2 \Delta t^2 \tau^{-2}}$ factor produced by the relative delay between optical pulses in Eq.(10b). In Fig.7c, the optimal conversion efficiency as a function of terahertz frequency is plotted along with numerical calculations at the corresponding points. Consistent with earlier depicted trends in Section 4.1 and 4.2, the conversion efficiency obtained via numerical calculations are smaller at larger terahertz frequencies due to a more adverse impact of the modification of phase-matching conditions by cascading effects.  Despite these limitations, conversion efficiencies of ~ 2% are predicted at 0.5 THz.  The peak terahertz fields are plotted in Fig.7d and corresponding numbers of cycles are plotted in Fig.7e. As with prior cases (Section 4.2, 4.3) ,

the peak electric fields obtained numerically are smaller than the analytic values at larger terahertz frequencies due to modification of phase-matching conditions by cascading effects. Focused field strengths of few 100 MV/m are observed. The obtained number of terahertz field cycles is similar for numerical and analytic calculations, consistent with the trends from Section 4.2-4.3. The reduction of interaction lengths due to cascading effects is evident in Fig.7f for larger frequencies as in other pulse sequence formats.

## 5. Practical considerations
*Self-focusing*

Since our studies were limited to one dimensional spatial calculations, we calculate self-focusing distances [42] as follows:

$$z_{sf} = \sqrt{\frac{\pi w_{in}^2}{\lambda_0} \frac{1}{2\pi n_{2,eff} In(\lambda_0)}} \tag{11}$$

Here $\lambda_0$ is the central wavelength of the optical pump, $n_{2,eff}$ is the effective nonlinear refractive index (Table.1), $w_{in}$ is the input pump beam radius and $I$ is the peak pump intensity. The peak intensity is calculated considering laser induced damage limitations from Eq.(7a) and is proportional to $\tau_d^{-1/2}$. $w_{in}$ is calculated assuming a total energy of 1 Joule in the pump pulse sequence and a flat-top beam profile. Due to damage fluence limitations from Eq.(7b), $w_{in}^2 = 1/(F_d \pi)$ is proportional to $\tau_d^{-1/2}$. From Eq.(11), the dependencies on $\tau_d$ of $I$ and $w_{in}$ compensate each other, resulting in $z_{sf}$ being independent of $\tau_d$ and equal to ~75 cm for 1 Joule pump pulse energies. This means that self-focusing should not pose an issue as long as operation is below the damage threshold. The caveat however is that $\tau_d$ has to be large enough to permit values of $w_{in}$ that can be accommodated within feasible crystals. Reduction of self-focusing distances would occur for beams with local spatial modulations. However, this issue could be circumvented with a suitable imaging system to clean up the beam prior to the PPLN.

*Diffraction of terahertz beams*

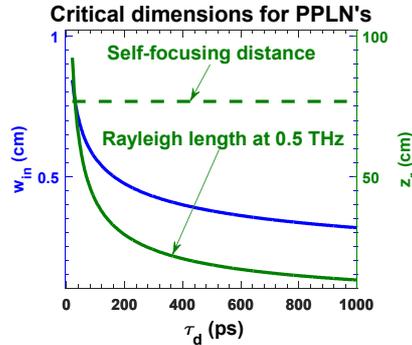

Fig.8. Self-focusing, terahertz Rayleigh lengths and requisite crystal apertures as a function of $\tau_{eff}$. Crystal apertures are within the grasp of current technology while self-focusing and diffraction lengths are longer than the optimal interaction lengths.

For an optical pump beam radius $w_{in}$, the terahertz beam radius would be $w_{in}/\sqrt{2}$ as a virtue of being generated by a second order nonlinear process. The Rayleigh length (as defined for a Gaussian beam) for a terahertz wave propagating inside the crystal is given by Eq.(12).

$$z_R = \frac{1}{2}\frac{\pi w_{in}^2 n_{THz} f_{THz}}{c} \tag{12}$$

In Fig.8, we plot the damage limited beam radii $w_{in}$, Rayleigh and self-focusing lengths at 0.5 THz as a function of the effective pulse duration $\tau_d$. It is seen that the Rayleigh lengths are significantly larger than the optimal interaction lengths required. For effective damage threshold durations $\tau_d$ = 100-500 ps, the maximum crystal apertures ($2w_{in} \times 2w_{in}$) required would be < 1.2 cm$^2$, which can be realized with demonstrated technology [31]

*Multi-photon absorption*
For pump wavelengths in the range of ~1μm, 4-photon absorption of the fundamental and 2-photon absorption of the second harmonic are possible. The former is expected to be weak even for intensities as large as 100 GWcm$^{-2}$ [21, 43], while the latter is not expected to be significant since the SHG is not phase-matched. However, for 800 nm, 3-photon absorption of the fundamental and 2-photon absorption of second harmonic+fundamental are possible [44]. These multi-photon processes are stronger than those for the case of ~1μm wavelengths. Therefore, for 800 nm, lower damage thresholds and conversion efficiencies are anticipated.

*Practical implementation of the pulse formats*
To generate highly efficient terahertz fields, three pulse formats – a burst of pulses, multiple quasi-CW lines and chirped and delayed broadband pulses have been presented. From a laser engineering point of view, the chirp and delay pulse format can be readily implemented with existing Titanium Sapphire lasers. It involves stretching of a pulse (which is typically done anyway during chirped pulse amplification), splitting and delaying it. However, high pulse energies from Titanium sapphire lasers are typically limited to low repetition rates. For higher average powers, Ytterbium based lasers [17] are strong candidates. However, the limited bandwidth of these lasers makes chirp and delay approaches less attractive. Therefore, a burst of pulses with a pulse separation of just a few picoseconds can be achieved by splitting and delaying one laser pulse. The multiple quasi-CW line approach relies on spatially superimposing multiple highly stable narrow band lasers. To achieve high pulse energies, modulators maybe needed prior to pulse amplification. Such multi quasi-CW line sources may also be self-generated starting with a single quasi-CW pump and weak quasi-CW optical seed pulse. The terahertz radiation, initially generated by the DFG between the pump and seed pulse then rapidly causes the optical spectrum to cascade, thus generating a series of quasi-CW lines [32].

## 6. Conclusion
Highly efficient terahertz generation approaches employing cascaded difference frequency generation (DFG) of pulse sequences in periodically poled lithium niobate (PPLN) were proposed and analyzed over a large parameter space. Optimal energy conversion efficiencies > 5%, peak electric fields of several hundred MV/m for terahertz pulses with hundreds of picoseconds durations are predicted using calculations employing pump depletion. At Joule level pumping, this translates to terahertz pulse energies >> 10 mJ. Analytic formulations for calculating the conversion efficiencies for arbitrary pulse formats were presented which included the effects of dispersion and absorption. These showed good qualitative and quantitative agreement with detailed numerical simulations including cascaded DFG, self-phase modulation, cascaded second harmonic generation and laser induced damage, particularly at low conversion efficiencies and terahertz frequencies. The physics of terahertz generation using pulse sequences in PPLN's was discussed. At sufficiently high conversion efficiencies, significant cascading of the optical spectrum results in a change in the phase-matching condition, thereby limiting conversion efficiency. Changing the PPLN period along the propagation direction and further mitigation of losses could lead to even higher conversion efficiencies >>10%.

**Acknowledgments**
The authors thank Prof. Erich.P.Ippen for helpful comments and Dr. Sergio Carbajo for initial laser induced damage threshold characterization. This work was supported by the European Research Council under the European Union's Seventh Framework Programme (FP/2007-2013) / ERC Grant Agreement n. 609920, the Center for Free-Electron Laser Science at DESY and the excellence cluster "The Hamburg Centre for Ultrafast Imaging- Structure, Dynamics and Control of Matter at the Atomic Scale" of the Deutsche Forschungsgemeinschaft.

## A. Analytic formulation

The Fourier transform for a function $f(t) = \int_{-\infty}^{\infty} f(\omega) e^{j\omega t} d\omega$. The corresponding inverse Fourier transform is defined as $f(\omega) = (2\pi)^{-1} \int_{-\infty}^{\infty} f(t) e^{-j\omega t} dt$. The definitions follow energy conservation (Parseval's theorem) according to $\int_{-\infty}^{\infty} |f(t)|^2 dt = 2\pi \int_{-\infty}^{\infty} |f(\omega)|^2 d\omega$.

*Derivation of Eq.(5)*
The nonlinear polarization term from Eq.(4) is substituted in Eq.(1). Setting $A_{THz}(\Omega, z) = \varphi(\Omega, z) e^{-\frac{\alpha(\Omega) z}{2}}$ reduces Eq.(1) to the following for $\omega_0 \gg 0$, where $\Delta\omega = \omega - \omega_0$ is the displacement from the central angular frequency of the pump.

$$\varphi(\Omega, z) = \varepsilon_0 \chi_{eff}^{(2)} \int_{-\infty}^{\infty} \int_0^z A_{op}(\Delta\omega + \Omega) A_{op}*(\Delta\omega) e^{-j\left(\Delta k + \beta'' \Delta\omega\Omega + \frac{\beta''\Omega^2}{2}\right)z} e^{\frac{\alpha z}{2}} dz d\Delta\omega.$$ For $\beta'' \Delta\omega\Omega \ll \alpha$,

we obtain: $A_{THz}(\Omega, z) = \varepsilon_0 \chi_{eff}^{(2)} \int_0^{\infty} A_{op}(\Delta\omega + \Omega) A_{op}*(\Delta\omega) \left[ \frac{e^{-j\left(\Delta k + \beta'' \Delta\omega\Omega + \frac{\beta''\Omega^2}{2}\right)z} - e^{-\frac{\alpha(\Omega) z}{2}}}{\alpha(\Omega) - 2j\Delta k} \right] d\Delta\omega$.

The integral, $\int_0^{\infty} A_{op}(\Delta\omega + \Omega) A_{op}*(\Delta\omega) d\Delta\omega$ may be written as $F\left(|A_{op}(t)|^2\right)$, upon expressing $A_{op}(\Delta\omega + \Omega) = (2\pi)^{-1} \int_{-\infty}^{\infty} A_{op}(t) e^{-j(\Delta\omega+\Omega)t} dt$ and reversing the order of integrals [45]. Proceeding along similar lines for other integrals, Eq.(5) is obtained.

## B. Simulation details
### B.1 Material properties
Since the terahertz frequency is much smaller than the optical frequency, the relevant second order nonlinear effect is the electro-optic effect. The electro-optic tensor element which maximizes terahertz generation in lithium niobate is $r_{33}$. This corresponds to extraordinarily polarized terahertz and optical fields. For lithium niobate, $r_{33} \sim 32$ pm/V. The effective nonlinear coefficient for bulk lithium niobate at a pump wavelength $\lambda_0$ is then $d_0 = r_{33} n(\lambda_0)^4 / 4$ [46]. At 1030 nm, $n(\lambda_0) = 2.15$, which yields a $d_0 = 168$ pm/V. The corresponding bulk susceptibility is then $\chi_0^{(2)} = 2d_0$. In PPLN structures, there is periodic inversion of the nonlinearity along the crystal direction $z$, which is described by $\chi^{(2)}(z) = \chi_0^{(2)} \text{sgn}\{\cos(2\pi\Lambda^{-1}z)\}$, where **sgn** is the signum function.

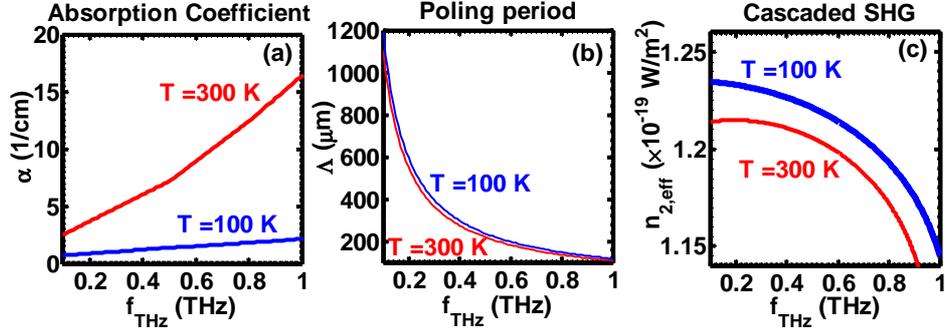

Fig.9(a) Terahertz absorption coefficients at various crystal temperatures.(b) PPLN periods for various phase-matched terahertz frequencies at different crystal temperatures (c) Values of $n_{2,eff}$ obtained using Eq.(13) at different crystal temperatures.

The terahertz absorption coefficients at 300 K and 100 K are obtained from [40] in Fig.9a. At 0.5 THz, the absorption coefficient at 300 K is 7.2 cm$^{-1}$. At 100K, the value reduces to about 1.4 cm$^{-1}$, representing a fivefold decrease. Even further decrease in absorption may be obtained by cooling the crystal to 10K, which yields an absorption coefficient of 0.25 cm$^{-1}$ at 0.5 THz [40]. In Fig.9b, the PPLN poling period $\Lambda$ for terahertz generation at various frequencies $f_{THz}$ is obtained in the phase-matched condition as $\Lambda = c / (f_{THz}[n_{THz}(\Omega) - n_g(\lambda_0)])$. The terahertz phase index at 0.5 THz is $n_{THz}(\Omega) \sim 4.96$ at 300 K and 4.73 at 100 K [39]. The group refractive index is $n_g(\lambda_0) = 2.21$ and is relatively insensitive to temperature change.

Second harmonic generation is highly phase-mismatched in bulk lithium niobate. Corresponding to the fundamental at 1030 nm with refractive index 2.15, the second harmonic at 515 nm has a refractive index $n_{SHG} = 2.24$. The corresponding phase-mismatch $\Delta k = 1.02 \times 10^6$ m$^{-1}$, which corresponds to a coherence length of $\pi \Delta k^{-1} \sim 3$ μm that is significantly smaller than the PPLN periods under consideration. However, phase-mismatched SHG can manifest itself as a third order nonlinearity which can considerably alter the self-phase-modulation effect. These cascaded SHG effects have an effective third order susceptibility given by the following [34].

$$\chi^{(3)}{}_{casc} = \sum_m -\frac{16\pi d_m^2}{3n_{SHG}\lambda_0}\frac{1}{\Delta k_m} \quad (13a)$$

$$\chi^{(3)} = \chi^{(3)}{}_{casc} + \chi^{(3)}{}_{bulk} \quad (13b)$$

Here, $d_m = 2d_{SHG}(m\pi)^{-1}\sin(m\pi/2)$ is the $m^{th}$ term in the Fourier series expansion of $d_{SHG}(z) = d_{SHG}\text{sgn}\{\cos(2\pi\Lambda^{-1}z)\}$ and $\Delta k_m = 4\pi\lambda_0^{-1}[n_{SHG} - n(\lambda_0)] - 2\pi m\Lambda^{-1}$ is the $m^{th}$ order phase-mismatch. The value of $d_{SHG} = 25$ pm/V for lithium niobate. The total third order susceptibility is given by Eq.(13b). $\chi^{(3)}{}_{bulk}$ includes instantaneous Kerr and Stimulated Raman Scattering (SRS) contributions and has a value of 6365 pm$^2$/V$^2$ [34]. The cascaded SHG contribution has a sign opposite of $\chi^{(3)}{}_{bulk}$ and amounts to $\chi^{(3)}{}_{casc} = -4428.8$ pm$^2$/V$^2$ in bulk LN (obtained by setting $d_m = d_{SHG}\delta_{m,0}$, where $\delta$ is the Dirac-Delta function). In bulk LN, this results in a total $\chi^{(3)} = 1936.2$ pm$^2$/V$^2$. The corresponding nonlinear refractive index coefficient $n_{2,eff} = 3\chi^{(3)}/4n(\lambda_0)^2\varepsilon_0 c$ is then equal to $1.17 \times 10^{-19}$ m$^2$/Watt, similar to reported values [47]. In Fig.9c, we plot the calculated $n_{2,eff}$ for PPLNs at various THz frequencies and crystal temperatures. Equation 13 is used for the calculations, with values of $m \neq 0$ ranging over negative and positive orders upto the 10$^{th}$ order.

**B.2 Numerical diagnostics**
In Fig. 10, we present the total energy (or fluence in the 1-D case) for the simulation in Section 3. Absorption is turned off to test the conservation of energy. As can be seen, energy is conserved very well over the simulated crystal lengths.

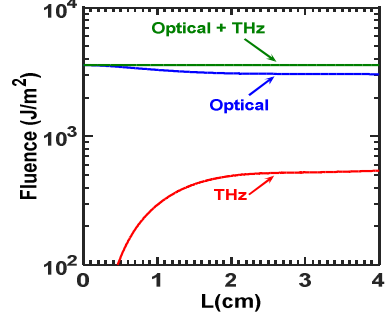

Fig.10. Verification of the numerical conservation of energy in the simulations.

**B.3 Pulse sequence formats**

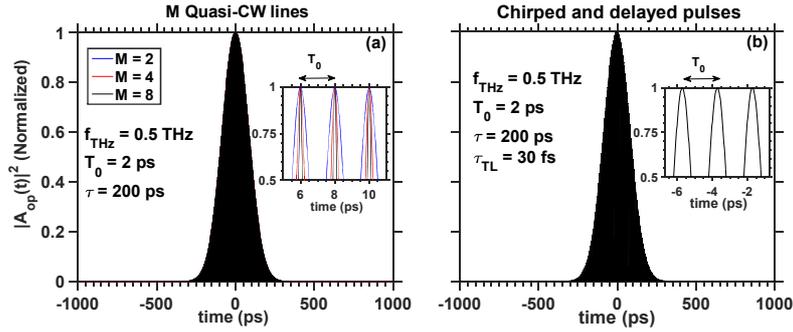

Fig.11(a). Temporal intensity profile corresponding to a spectrum with $M$ quasi-CW lines, each corresponding to a FWHM duration $\tau$ =200 ps and separated by the terahertz frequency $f_{THz}$=0.5 THz. A sequence of 'sub-pulses' in time, separated by the time period of the terahertz wave $T_0$=2 ps is obtained. The duration of sub-pulses reduces with increasing $M$. (b) Temporal intensity profile corresponding to the overlap of two broadband pulses with transform limited duration $\tau_{TL}$=30fs, chirped to $\tau$ =200ps. A sequence of sub-pulses separated by $T_0$=2 ps is obtained. The case is similar to that of $M$=2 quasi-CW lines.

Consider a pair of quasi-CW lines, separated by the desired terahertz frequency $f_{THz}$. Each quasi-CW line has a linewidth corresponding to a transform limited full-width at half-maximum (FWHM) duration $\tau$. In the time domain, such a spectrum corresponds to an intensity $\left|A_{op}(t)\right|^2$ that is proportional to $e^{-\frac{4\ln 2 t^2}{\tau^2}}\left|e^{-j2\pi f_{THz}t}+1\right|^2$ or $e^{-\frac{4\ln 2 t^2}{\tau^2}}\cos^2(\pi f_{THz}t)$. The corresponding intensity profile is a Gaussian envelope with FWHM duration $\tau$, containing a sinusoidal modulation which appears as a sequence of 'sub-pulses' separated by the time period of the terahertz wave $T_0 = f_{THz}^{-1}$ in time. The FWHM duration of each sub-pulse is obtained by setting $\cos^2(\pi f_{THz}t)=1/2$. This yields sub-pulse FWHM duration of $T_0/2$. The total number of pulses in an envelope with FWHM duration $\tau$ is $N = \tau/T_0$. The effective damage threshold duration $\tau_d$ which is $N$ times the individual sub-pulse duration is then $\tau_d = N T_0/2 = (T_0^{-1}\tau)T_0/2 = \tau/2$. If there are $M$ lines instead of just 2, the envelope retains a FWHM duration $\tau$ but the additional bandwidth results in proportionally shorter 'sub-pulse' durations (Fig.11a). Therefore, $\tau_d = \tau/M$ for arbitrary $M$.

Similarly, when two broadband chirped pulses are overlapped with a suitable delay, $|A_{op}(t)|^2$ becomes proportional to $e^{-\frac{4\ln 2 t^2}{\tau^2}}\cos^2(bt\Delta t + \varphi)$ which is also a Gaussian envelope with sinusoidal modulations (See Fig.11b). Since $2b\Delta t = 2\pi f_{THz}$, this also appears as a sequence of 'sub-pulses' separated by $T_0$ with a sub-pulse duration $T_0/2$. Therefore, $\tau_d = \tau/2$ for chirp and delay approaches.